\documentclass[aps,prl,showpacs,twocolumn,groupedaddress]{revtex4} 
\usepackage[latin1]{inputenc} 
\usepackage{graphicx,epsfig,epstopdf} 
 
\newcommand{\hpiplong}{(C$_{5}$H$_{12}$N)$_2$CuBr$_4$} 
\newcommand{\hpip}{(Hpip)$_2$CuBr$_4$} 

\begin{document} 
 
\title{Thermodynamics of the Spin Luttinger Liquid in a Model Ladder Material} 
 
\author{ 
Ch. R\"uegg,$^{1}$ 
K. Kiefer,$^{2}$ 
B. Thielemann,$^{3}$ 
D. F. McMorrow,$^{1}$ 
V. Zapf,$^{4}$ 
B. Normand,$^{5}$ 
M. B. Zvonarev,$^{6}$ \\ 
P. Bouillot,$^{6}$ 
C. Kollath,$^{7}$ 
T. Giamarchi,$^{6}$ 
S. Capponi,$^{8,9}$ 
D. Poilblanc,$^{9,8}$ 
D. Biner,$^{10}$ 
and K. W. Kr\"amer$^{10}$} 
 
\affiliation{ 
$^1$London Centre for Nanotechnology; 
University College London; London WC1E 6BT; United Kingdom\\ 
$^2$BENSC; Helmholtz Centre Berlin for Materials and Energy; D--14109 Berlin; 
Germany\\ 
$^3$Laboratory for Neutron Scattering; ETH Zurich and Paul Scherrer 
Institute; CH--5232 Villigen PSI; Switzerland\\ 
$^4$NHMFL; Los Alamos National Laboratory; Los Alamos; New Mexico 87545; 
USA\\ 
$^5$Institute for Theoretical Physics; Ecole Polytechnique F\'ed\'erale 
de Lausanne; CH--1015 Lausanne; Switzerland\\ 
$^6$DPMC-MaNEP; Universit\'e de Geneve; CH-1211 Geneva 4; Switzerland\\ 
$^7$Centre de Physique Th\'eorique; \'Ecole Polytechnique; CNRS; 91128 
Palaiseau; France\\ 
$^8$Universit\'e de Toulouse; UPS; Laboratoire de Physique Th\'eorique; IRSAMC; 
F--31062 Toulouse; France\\ 
$^9$CNRS; UMR 5152; F-31062 Toulouse; France\\
$^{10}$Department of Chemistry and Biochemistry; University of Bern; 
CH-3000 Bern 9; Switzerland \\} 
 
\date{\today} 
 
\begin{abstract} 
The phase diagram in temperature and magnetic field of the metal--organic, 
two--leg, spin--ladder compound (C$_{5}$H$_{12}$N)$_{2}$CuBr$_{4}$ is 
studied by measurements of the specific heat and the magnetocaloric effect. 
We demonstrate the presence of an extended spin Luttinger--liquid phase 
between two field--induced quantum critical points and over a broad range 
of temperature. Based on an ideal spin--ladder Hamiltonian, comprehensive 
numerical modelling of the ladder specific heat yields excellent 
quantitative agreement with the experimental data across the complete 
phase diagram. 
\end{abstract} 
 
 
\maketitle 
 
Quantum spin systems display a remarkable diversity of fascinating physical 
behavior. This is especially true for systems such as spin ladders, which 
have a gapped or a gapless ground state, respectively, for an even or an 
odd number of ladder legs \cite{Dagotto95}. For two--leg ladders, and in 
general for any even leg number, quantum phase transitions (QPTs) between 
gapped and gapless phases can be driven by an external magnetic field. 
While these QPTs are generic in quantum magnets \cite{Giamarchi08}, the 
nature of the gapless phase depends crucially on the dimensionality of the 
spin system. In two and higher dimensions, a quantum critical point (QCP) 
separates the low--field, quantum disordered (QD) phase, with gapped triplet 
excitations, from a gapless phase with long--range antiferromagnetic (AF) 
order, which can be well described as a Bose--Einstein Condensate (BEC) of 
magnons \cite{Giamarchi08,Giamarchi99,nikuni00_tlcucl3_bec}.

By contrast, for one--dimensional (1D) systems such as ladders, both 
long--ranged magnetic order and BEC 
are precluded due to phase fluctuations. In addition, spin excitations 
are best viewed as interacting fermions, whereas a bosonic representation 
pertains in higher dimensions.
The physics of the gapless phase in 1D is thus quite different. It is 
a (spin) Luttinger liquid (LL) \cite{giamarchi_book_1d}, 
and is a key component of the rich phase diagram presented in Fig.~1 
\cite{Giamarchi99,Sachdev94,furusaki_correlations_ladder,Wang,Wessel}. 
In the LL, the spectrum is gapless with algebraically
decaying spin correlations. Because there is no finite order parameter, 
the LL regime 
is reached from the high--temperature, classical regime through a crossover 
rather than a phase 
transition. Nevertheless, clear manifestations of LL behavior are expected 
not only in the correlation functions 
but also in thermodynamic quantities such as the magnetization and 
specific heat. 

\begin{figure}[t!] 
\begin{center} 
\includegraphics[width=0.40\textwidth]{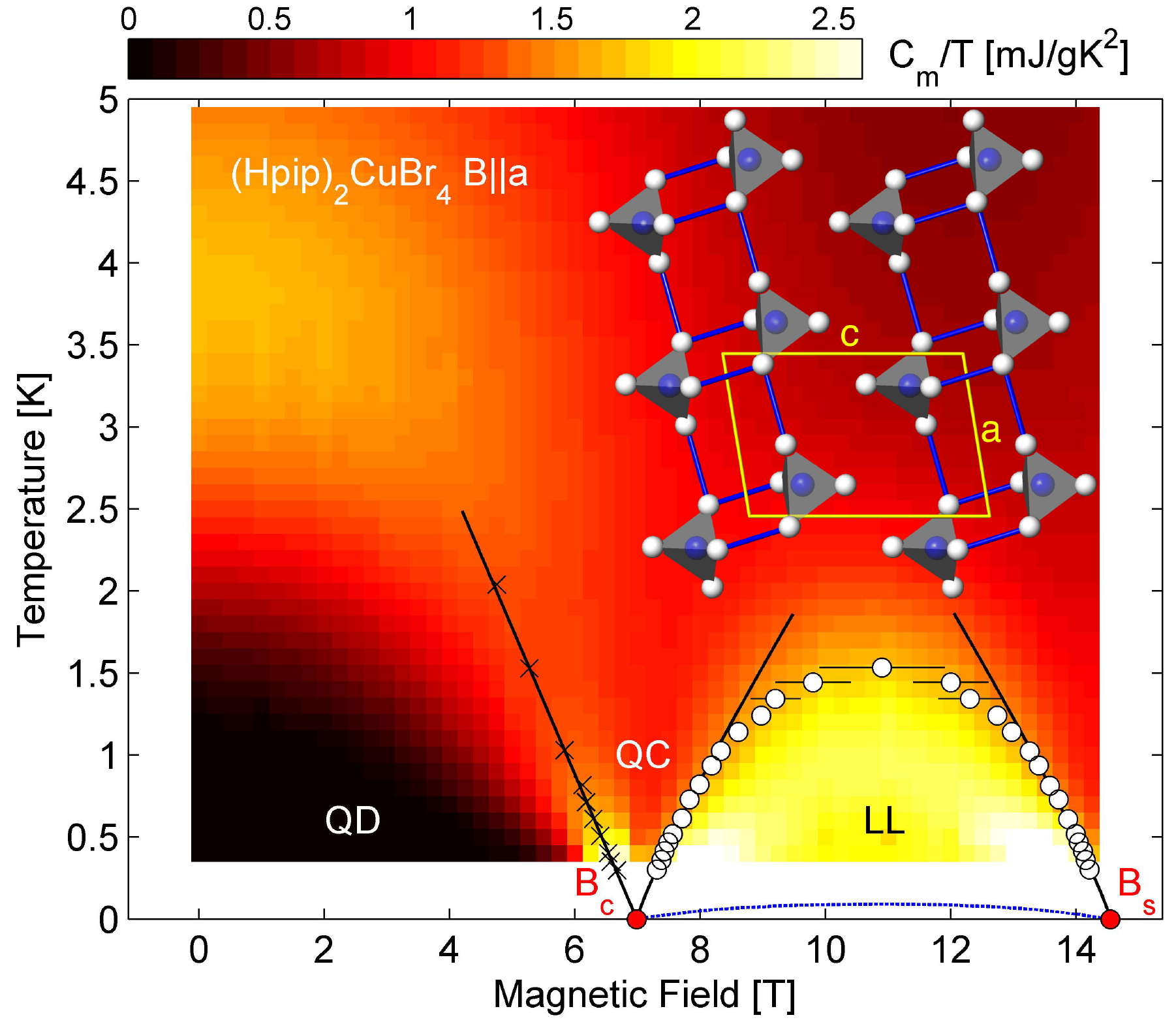} 
\end{center} 
\caption{\small Field--temperature phase diagram of the spin--ladder 
compound \hpip, showing quantum disordered (QD), quantum critical 
(QC), and spin Luttinger liquid (LL) phases. QCPs occur at $B_c$ 
(closing of spin triplet gap $\Delta$) and $B_s$ (spin system fully 
polarized). The contour plot shows the magnetic specific heat as 
$C_{m}(T,B)/T$. Local maxima from the reduction of the triplet gap 
by the Zeeman effect are indicated by crosses. Circles 
denote the LL crossover based on measurements of the magnetocaloric 
effect [Fig.~4], black lines are fits to extract the critical fields, 
and the dashed blue line indicates the onset of long--ranged order below 
100 mK \cite{Berthier,Thielemann}. Inset: lattice structure of \hpip~in 
projection along the $b$--axis, with Cu atoms blue and Br white. } 
\label{fig1} 
\end{figure} 

However, materials in which to explore such effects are rather rare. 
Investigations of the spin excitations and thermodynamic properties of 
ladder compounds have to date been performed primarily on copper oxides.
In these materials, the exchange interactions are typically some hundreds 
of meV, and thus the phases are not easily controlled by a magnetic field. 
Candidate 1D materials in which thermodynamic properties have been 
studied around the QPT include the bond--alternating $S=1/2$ and $S=1$ 
chains F$_5$PNN \cite{Yoshida} and NTEMP 
\cite{Hagiwara}, the $S=1$ Haldane system NDMAP \cite{Honda, Chen}, 
and the $S=1/2$ system CuHpCl \cite{Chaboussant}, which was for some 
time considered to be a spin ladder \cite{Stone}. While measurements 
in these materials show indications of LL behavior in parts of the 
field--temperature phase diagram, their magnetic properties are 
influenced in large part by the presence of significant single--ion 
anisotropy and/or three--dimensional (3D) interactions between the 
chains, which tend to dominate the low--temperature specific heat at all 
fields \cite{DM}. Hence we have pursued the search for materials suitable 
to study the intrinsic spin LL physics by seeking those showing, at 
minimum, a clear separation of energy (temperature) scales between 
1D and 3D interactions.
 
Here we present the results of thermodynamic measurements on an 
exceptional two--leg ladder material, the compound 
piperidinium copper bromide \hpiplong~\cite{Patyal,Watson,Lorenz,Anfuso, 
Berthier,Thielemann}, where all of the phases of interest can be accessed,
as summarized in Fig.~1. In particular, we find an extended region 
in the phase diagram, between 0.1 K and 1.5 K, 
where a spin LL is observed. We show that the crossover into the LL 
is signaled by clear features in both the specific heat and the magnetization.
In the gapless spin LL, the magnetic 
specific heat is linear at low temperatures. Its field-- and 
temperature--dependence are in excellent agreement with numerical 
calculations involving no free parameters. This demonstrates that the 
material is very accurately described by a minimal spin--ladder 
Hamiltonian. 
 
High--quality single crystals of \hpiplong, abbreviated \hpip~in the 
following, were 
grown from solution. In this material, the $S=1/2$ magnetic moments of 
the Cu$^{2+}$ ions are arranged in a ladder--like structure along the 
$a$--axis [Fig.~1, inset]. The rungs ($J_{r}$) of this ladder are formed 
by two equivalent Cu--Br--Br--Cu superexchange paths with a center of 
inversion symmetry \cite{DM}, while the legs ($J_{l}$) involve one similar 
but longer interaction path. The ladder units (Cu$_{2}$Br$_{8}$)$^{4-}$ 
are well separated by the organic (C$_{5}$H$_{12}$N)$^{+}$ cations, 
which contribute only very little to the electronic properties of the 
host structure, and hence any magnetic exchange between ladders ($J'$) 
is expected to be small. Direct measurements of these interactions, 
$J_{r}$=12.9(2) K, $J_{l}$=3.3(3) K, and $J'<$ 100 mK, based on inelastic 
neutron scattering experiments \cite{Thielemann_INS}, are in very 
close agreement with the values extracted from magnetostriction \cite{Anfuso} 
and nuclear magnetic resonance (NMR) \cite{Berthier} measurements. 
 
The specific heat and magnetocaloric effect (MCE) were measured on a 
purpose--built calorimeter at the Helmholtz Centre Berlin (Laboratory for 
Magnetic Measurements at BENSC), using single 
crystals of masses 4.78 mg and 9.69 mg in the respective temperature and 
field ranges 0.3 K to 15 K and 0 T to 14.5 T. The field was applied 
parallel to the crystallographic $a$--axis, a geometry in which we 
obtained the values $B_{c}=6.99(5)$ T and $B_{s}=14.4(1)$ T for the 
two QCPs [Fig.~1]. The specific heat was extracted from a quasi--adiabatic 
relaxation technique and, using the same set--up, the MCE was recorded with 
a sweep rate of 0.05 T per minute.

\begin{figure}[t!] 
\begin{center} 
\includegraphics[width=0.45\textwidth]{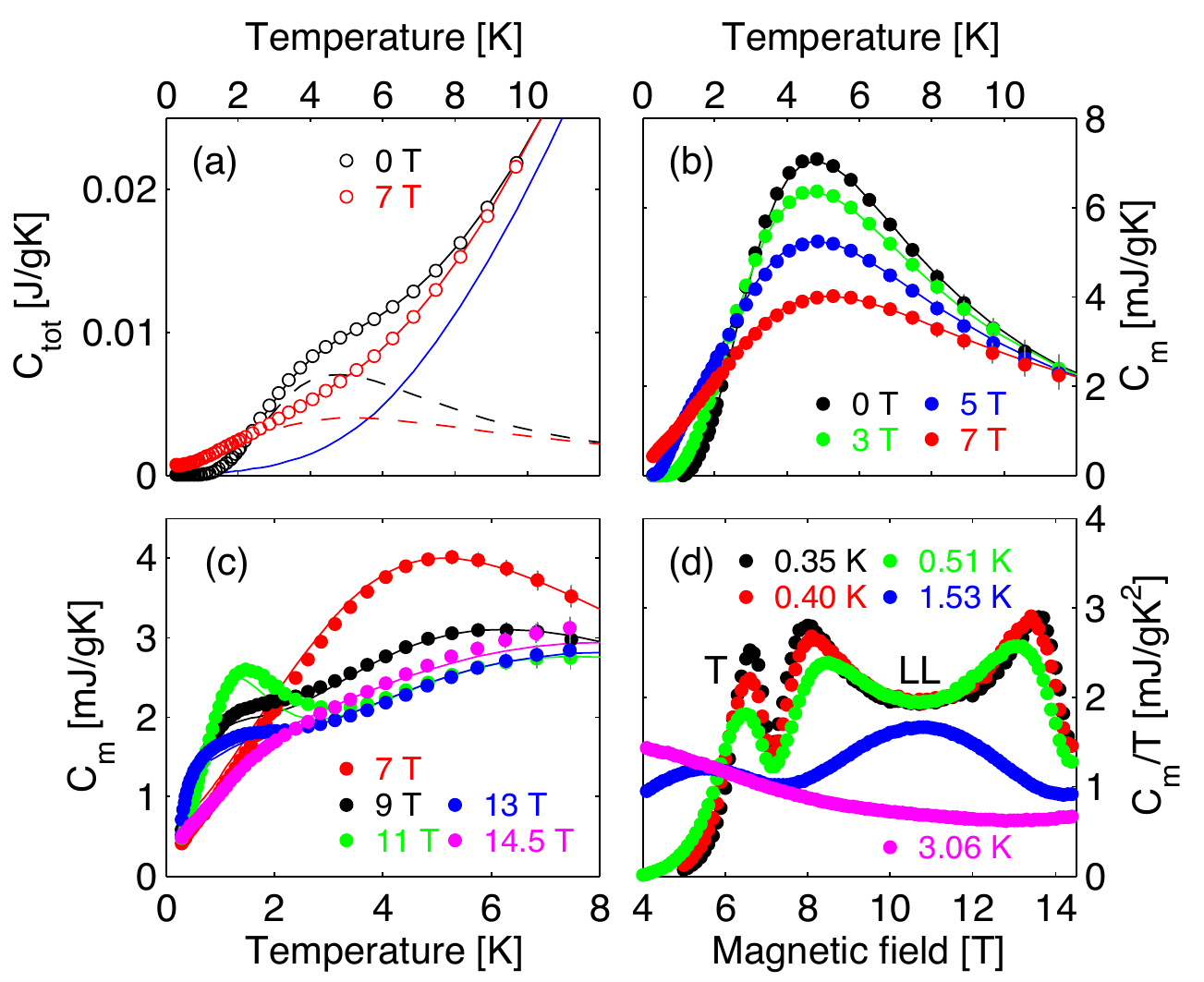} 
\end{center} 
\caption{\small (a) Measured total specific heat $C_{tot}(T)$ at 
fixed magnetic field. Solid (dashed) lines are based on fits to 
the data before (after) background subtraction. The blue line indicates 
the uniform, non--magnetic background. (b) Magnetic specific heat $C_{m} 
(T,B)$ for $B \le B_c$, and (c) $B_c \le B \le B_s$. Lines in (a--c) 
show $C_{m}^{ED}$, and are based on ED and DMRG calculations, as 
explained in the text. (d) $C_{m}(T,B)/T$ measured at fixed temperature. 
The region with linear temperature--dependence of the specific heat is 
indicated by LL, while T marks the peak due to the softening triplet.} 
\label{fig2} 
\end{figure} 
 
In Fig.~1 the magnetic component of the specific heat $C_{m}/T$ of 
\hpip~is presented across the entire phase diagram. It shows clearly 
three distinct regimes: QD, quantum critical (QC), and spin LL. The 
contour plot was obtained from 27 scans in field and temperature, after 
subtraction of a field--independent lattice contribution $C_{l}(T)$ and 
of a small nuclear term, which both are determined from a simultaneous 
fit to all available data. 

In Fig.~2 (a)--(d) we show individual measurements of the total and 
the magnetic specific heat, respectively $C_{tot}(T)$ and $C_{m}(T,B)$. 
In the QD regime, $B\le B_{c}$, $C_{m}$ shows a single peak at approximately 
5 K [Fig.~\ref{fig2}(b)]. This peak is attributable to the triplet 
excitations of the ladder, and is 
exponentially activated at lower temperatures due to the presence of 
the spin gap $\Delta$ \cite{Troyer}. With increasing field, the gap is 
reduced by the Zeeman effect ($\Delta\rightarrow\Delta-g\mu_{B} B$). Field 
scans such as those in Fig.~2(d) show most clearly the reduction of the 
gap and are used to extract the critical field $B_{c}$ in Fig.~1, yielding 
very good agreement with determinations by complementary experimental 
techniques \cite{Watson,Lorenz,Anfuso,Berthier,Thielemann}. 
 
The specific heat changes dramatically for fields $B>B_{c}$, which we 
explain by the formation of the LL phase [Fig.~2(c)]. While at high 
temperature $C_{m}$ is dominated by the (gapped) $S_z=0$ triplet states, 
at low temperature an additional peak develops. Below this peak, the 
temperature--dependence remains linear up to $B = B_s\approx14.5$ T,
with a field--dependent slope. The linearity of $C_{m}$ is demonstrated 
in Fig.~\ref{fig2}(d). For fields near the maximum of the 
LL dome, the ratio $C_{m}/T$ measured at different temperatures 
collapses onto the same curve. This temperature--dependence is consistent
with the presence of gapless spinon excitations with a finite velocity 
$u$, its slope being inversely proportional to $u$ \cite{giamarchi_book_1d}. 
The first peak thus occurs when the temperature is large enough to probe 
deviations from this linear regime. It can be taken as an estimate of 
the crossover to enter the LL, and is visible in Fig.~\ref{fig1} for 
$B_{c}<B<B_{s}$. The field--dependence of $C_{m}$ is almost symmetric 
about $(B_{c}+B_{s})/2\approx10.7$ T. In the strong--coupling limit, 
$J_r/J_l \gg 1$, a perfect symmetry would be expected due to the exact 
particle--hole symmetry of the XXZ chain in a field 
\cite{Chaboussant,Strong_Coupling}. Here we observe clearly deviations 
characteristic of the underlying ladder structure. Similar effects are 
also visible in spin correlation functions and in the low--temperature 
phase diagram, which can be measured by NMR \cite{Berthier} and neutron 
scattering \cite{Thielemann}. 
  
At $B>B_s$, the specific heat becomes exponentially activated again due 
to the opening of a field--dependent spin gap in the fully saturated phase. 
However, this regime is close to the limit of our experimental window, and 
so the high--field phase is not investigated further here. 

The experimental data have been compared with several theoretical 
calculations, and the agreement is remarkable (Fig.~2). Numerical results 
were obtained by exact diagonalization (ED) and by adaptive, time--dependent 
density--matrix renormalization--group (DMRG) calculations \cite{DMRG_1}, 
both performed for a single ladder with $J_{r}$=13 K, $J_{r}/J_{l}$=4, 
and $g$=2.06 ({\it i.e.}~no free parameters). We stress that in both 
techniques it is important to retain a sufficient number of ladder states
for a quantitative description of thermodynamic data. The DMRG 
calculations (2$\times$40 spins) may be regarded as the definitive 
behavior of this model. In the ED calculations, the specific heat of 
even-- (odd--)length ladders converges rapidly from above (below) to 
the infinite--size limit; thus finite--size effects are essentially 
removed here by taking an average between systems of 2$\times$10 and
2$\times$11 spins. The ED and DMRG results are indistinguishable both in the 
QD phase [Fig.~2(b)] and in the LL regime [Fig.~2(c)]. Slight deviations 
from the experimental data are found only close to the upper critical field 
$B_s$ and at $11$T.

\begin{figure} 
\begin{center} 
\includegraphics[width=0.45\textwidth]{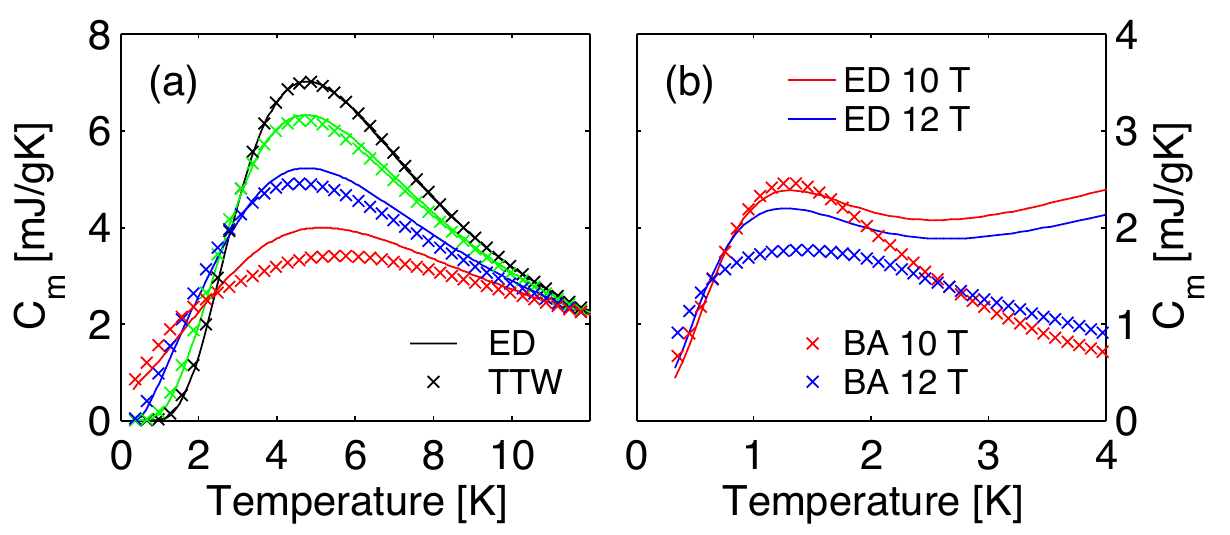} 
\end{center} 
\caption{\small Comparison of calculated ladder specific heat. (a) 
ED vs. TTW for magnetic--field values $B \le B_c$ [Fig.~2(b)]. (b) 
ED vs. BA for some magnetic--field values $B > B_c$ [Fig.~2(c)].} 
\label{fig3} 
\end{figure} 

Some physical insight into the numerical results is afforded by 
two approximate treatments. A statistical Ansatz (TTW) \cite{Troyer} 
developed for spin ladders, and shown previously to describe very 
accurately the thermal renormalization of triplet excitations in the 
3D dimer system TlCuCl$_3$ \cite{Ruegg05}, uses the correct number of 
states but applies their hard--core constraint only globally. In the spin 
ladder, this approximation underestimates the local energy of the excited
states, leading to a systematic shift of weight to lower energies as 
$B \rightarrow B_c$ [Fig.~\ref{fig3}(a)]. A mapping of the lowest two 
modes of the ladder Hamiltonian onto an effective $S=1/2$ XXZ chain 
\cite{Chaboussant,Strong_Coupling}, whence thermodynamic quantities 
are computed exactly from the Bethe Ansatz (BA), is very accurate for 
the low--energy physics at $B>B_c$, but cannot reproduce the 
heat capacity at higher temperatures because of the missing triplet 
states ($S_z=0,-1$) [Fig.~\ref{fig3}(b)]. We conclude that the 
thermodynamic properties of \hpip~are described very accurately by a 
model of a single two--leg ladder, and that comprehensive measurements 
of the specific heat identify an extended LL regime.

\begin{figure}[t!] 
\begin{center} 
\includegraphics[width=0.45\textwidth]{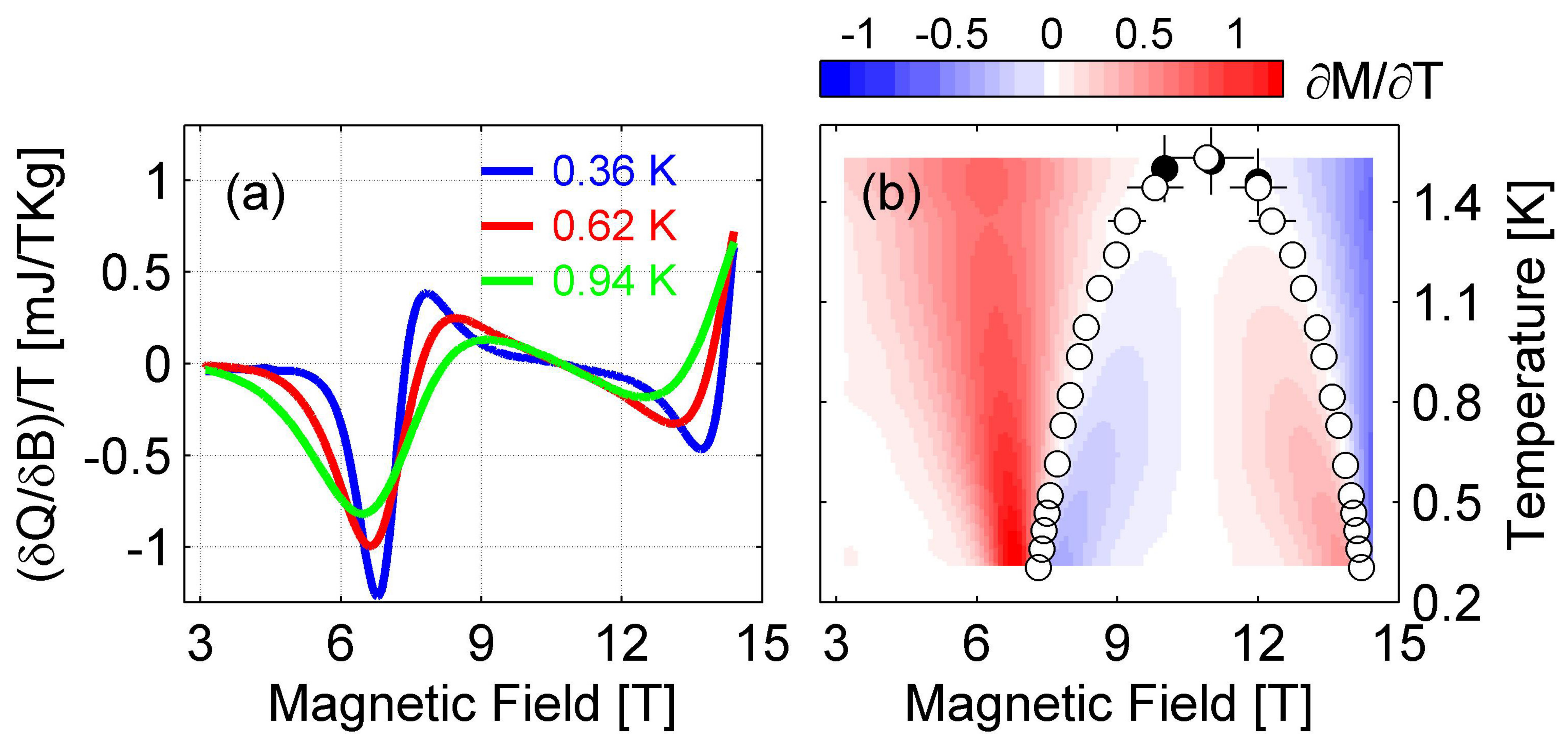} 
\end{center} 
\caption{\small Magnetocaloric effect in \hpip. (a) Heat--flow $\delta Q$ 
to and from 
the sample as a function of magnetic field divided by temperature, 
$(\delta Q/\delta B)/T$. (b) Contour plot of $\partial M/ \partial T$ as a 
function of field and 
temperature. White circles denote the phase boundary derived from 
$\partial M/ \partial T=0$ (see also Fig.~1), while black circles are 
maxima in the specific heat, 
$\partial C_{m}/\partial T=0$, obtained at fixed field.} 
\label{fig4} 
\end{figure}

We turn now to a different observable, the uniform magnetization ($M$), 
which is notoriously difficult to measure at temperatures below 1.5 K. 
Very precise measurements can be obtained by NMR \cite{Berthier}, but 
here we use an alternative method to probe the crossover to the LL. 
We determine the derivative of the magnetization with 
respect to temperature using the relation $(\delta Q/\delta B)/T$
 = $-(\partial M/\partial T)|_{B}$, where $\delta Q$ is the amount 
of heat created or absorbed by the sample for a field change $\delta B$ 
due to the MCE. Figure \ref{fig4} shows both representative 
$(\delta Q/\delta B)/T$--traces (corrected for a small base--line 
drift at higher temperatures) and a contour plot of all available data, 
presenting directly $\partial M/ \partial T$. In the free--fermion model, 
which is an excellent qualitative description of spins near the QCP in 1D 
\cite{giamarchi_book_1d}, 
and in more refined approaches \cite{Wang,Wessel,Oshikawa}, the magnetization
presents a minimum or  maximum as a function of temperature ($\partial M/ 
\partial T=0$). These extrema occur when the temperature matches the 
chemical potential, and thus provide another determination of the crossover 
temperature for the LL phase. The extracted phase boundary and the positions 
of the peaks in the specific heat agree well within expectations, as 
demonstrated by the solid symbols in Fig.~\ref{fig4}(b).

The structures present here in the magnetization and specific heat differ 
markedly from those occurring when there is a BEC. In that case, because 
a real phase transition occurs, the specific heat diverges with a 
$\lambda$--type anomaly. Such a shape has been observed at BEC transitions 
in higher--dimensional materials such as BaCuSi$_2$O$_6$ \cite{Jaime}. At the 
same temperature the magnetization develops a minimum, but with a cusp--like 
structure \cite{Giamarchi99}, which has been observed in TlCuCl$_3$ 
\cite{nikuni00_tlcucl3_bec}. For 1D spin ladders, the magnetization minimum 
and the specific--heat peak have a different origin: they correspond 
to the crossover to the LL regime. Thus there is no divergence in the 
specific heat and the magnetization minimum is analytic, reflecting the 
absence of a phase transition; the temperatures of the two features, 
although similar, are not identical. For \hpip,
a real phase transition of BEC type does occur at a much lower temperature, 
$T_N \approx 100$ mK (Fig.~1, \cite{Berthier,Thielemann}),
due to a 3D coupling of the ladders.

In summary, we have measured the specific heat and magnetocaloric 
effect in the metal--organic, two--leg spin ladder \hpiplong. The 
excellent low--dimensionality and optimal energy scale of the exchange 
interactions make this material unique, and allow a detailed 
investigation of the phase diagram in temperature and in fields up to 
magnetic saturation for the quantum spin ladder. We find an extended 
region of spin Luttinger liquid behavior over at least one order of 
magnitude in temperature, lying clearly above any three--dimensional 
physics triggered by residual interladder interactions. The 
high--precision experimental data have been analyzed using
the most advanced exact diagonalization and density matrix 
renormalization group techniques to calculate thermodynamic quantities 
for all of the phases ({\it i.e.}~across two quantum
critical points). From the direct and parameter--free fit of the 
experimental and numerical 
results, we conclude that \hpiplong~is remarkably well described 
by a minimal spin--ladder Hamiltonian, with other possible effects 
(frustrated interactions, Dzyaloshinskii--Moriya terms, lattice coupling) 
being very small. Hence the material offers unprecedented opportunities 
to investigate the intrinsic physics of low--dimensional quantum systems. 
 
We are grateful to C. Berthier and F. Essler for helpful discussions. 
This work was supported by the Royal Society, EPSRC, the Wolfson Foundation,
the network 'Triangle de la Physique', the Swiss National Science Foundation
through the NCCR MaNEP and Division II, and the French National Council (ANR).
S.C. thanks Calmip (Toulouse) for computing time.

\end{document}